\documentclass[doublecol]{epl2}

\pdfoutput=1

\title{Synthesis, crystal structure and spin-density-wave anomaly of the iron arsenide-fluoride \SFA}
\shorttitle{Synthesis, structure and SDW anomaly of the iron arsenide fluoride \SFA}

\author{Marcus Tegel\inst{1} \and Sebastian Johansson\inst{1} \and Veronika Wei\ss\inst{1} \and Inga Schellenberg\inst{2} \and Wilfried Hermes\inst{2} \and Rainer P\"{o}ttgen\inst{2} \and Dirk Johrendt\inst{1}}
\shortauthor{M. Tegel \etal}

\institute{
  \inst{1} Department Chemie und Biochemie, Ludwig-Maximilians-Universit\"{a}t M\"{u}nchen, Butenandtstrasse 5-13 (Haus D), 81377 M\"{u}nchen, Germany\\
  \inst{2} Institut f\"{u}r Anorganische und Analytische Chemie, Universit\"{a}t M\"{u}nster, Corrensstrasse 30,
D-48149 M\"{u}nster, Germany
}

\abstract{
The new quaternary iron arsenide-fluoride \SFA~with the tetragonal ZrCuSiAs-type structure was synthesized and the crystal structure was determined by X-ray powder diffraction ($P4/nmm$, a = 399.30(1), c = 895.46(1)~pm). \SFA~undergoes a structural and magnetic phase transition at 175~K, accompanied by strong anomalies in the specific heat, electrical resistance and magnetic susceptibility. In the course of this transition, the space group symmetry changes from tetragonal ($P4/nmm$) to orthorhombic ($Cmme$). $^{57}$Fe M\"{o}ssbauer spectroscopy experiments show a single signal at room temperature at an isomer shift of 0.30(1)~$mm/s$ and magnetic hyperfine-field splitting below the phase transition temperature. Our results clearly show that \SFA~exhibits a spin density wave (SDW) anomaly at 175~K very similar to LaFeAsO, the parent compound of the iron arsenide-oxide superconductors and thus \SFA~may serve as a further parent compound for oxygen-free iron arsenide superconductors.}

\begin{document}

\newcommand{\SFA}{SrFeAsF}
\newcommand{\CFA}{CaFeAsF}
\newcommand{\BaFA}{BaFe$_{2}$As$_{2}$}
\newcommand{\SrFA}{SrFe$_{2}$As$_{2}$}

\maketitle

\section{Introduction}

\begin{figure}[h]
\center{
\includegraphics[width=50mm]{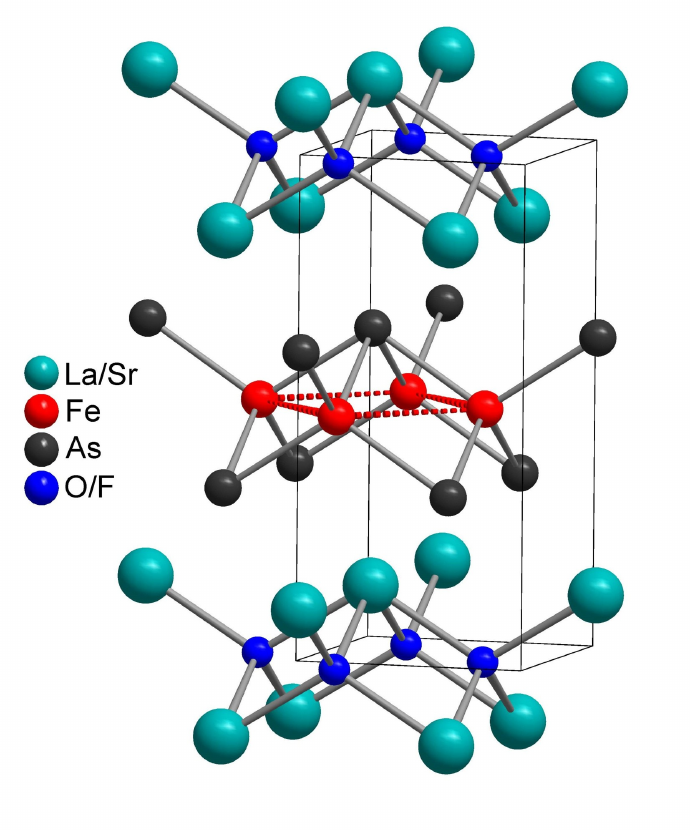}
\caption{Crystal structure of LaFeAsO and SrFeAsF.}
\label{fig:Structure}
}
\end{figure}

The discovery of superconductivity in doped iron-arsenides with ZrCuSiAs-type \cite{Hosono-2008} and ThCr$_2$Si$_2$-type \cite{Rotter-2-2008} structures with critical temperatures up to 55 K \cite{Ren-55K, Angew-2008} have heralded a new age in superconductivity research. The parent compounds of these superconductors are built up by alternating layers of (FeAs)$^-$ and (LaO)$^+$ or $\frac{1}{2}$Ba$^{2+}$, respectively. The crystal structure of LaFeAsO is depicted in Figure~\ref{fig:Structure}. Structural phase transitions from tetragonal to orthorhombic lattice symmetry occur prior to antiferromagnetic ordering \cite{Cruz-2008, Rotter-1-2008} and both can be suppressed by electron or hole doping, which also induces superconductivity. At first it was believed that this suppression is a prerequisite to superconductivity, but recent results strongly suggest that superconductivity coexists with the orthorhombic and magnetic phases for low doping at least in SmFeAsO and \BaFA \cite{Margadonna-2008, Rotter-3-2008, Chen-BKFA-2008}. Beyond strong efforts to shed light on the properties and mechanism of superconductivity, the search for further materials with possibly higher critical temperatures is still going on.

In this letter, we report on the iron arsenide fluoride SrFeAsF with tetragonal ZrCuSiAs-type structure, where the (LaO)$^+$ layers of LaFeAsO are replaced for (SrF)$^+$ layers. Our results clearly reveal that SrFeAsF exhibits a spin-density wave anomaly at 175 K as typical for these compounds and therefore may serve as a new parent compound for high-$T_c$ superconductors.

%
%
\section{Experimental}

\SFA~was synthesized by heating a mixture of strontium fluoride, distilled strontium metal, H$_2$-reduced iron-powder and sublimed arsenic at a ratio of 1:1:2:2 in a sealed niobium tube under an atmosphere of purified argon. The mixture was heated to 1173~K at a rate of 50~K/h, kept at this temperature for 40 h and cooled down to room temperature. The product was homogenized in an agate mortar, pressed into pellets and sintered at 1273~K for 48 h. The obtained black crystalline powder of \SFA~is stable in air.


%
%
Phase purity was checked by X-ray powder diffraction on a Huber G670 Guinier imaging plate diffractometer (Cu-$K_{\alpha_{1}}$ radiation, Ge-111 monochromator), equipped with a closed-cycle cryostat. Rietveld refinements of \SFA~were performed with the GSAS package \cite{GSAS} using Thompson-Cox-Hastings functions with asymmetry corrections as reflection profiles. \cite{Finger-Cox-Jephcoat} To obtain lattice parameter changes of highest possible accuracy, asymmetry parameters as well as the gaussian parameters $U$ and $V$ were refined at 10~K and held constant for all other temperatures.

A $^{57}$Co/Rh source was available for $^{57}$Fe M\"{o}ssbauer spectroscopy investigations. The \SFA~sample was placed in a thin-walled PVC container. The measurement was run in the usual transmission geometry at temperatures between 4.2 and 298~K. The source was kept at room temperature. The total counting time was approximately one day per spectrum. Fitting of the spectra was performed with the NORMOS-90 program system.\cite{Moessbauer_Normos}

Heat capacity measurements between 2.1 and 305 K were carried out on a Quantum Design Physical Property Measurement System (PPMS). The sample was fixed to the platform of a pre-calibrated heat capacity puck using Apiezon N grease. The magnetic susceptibility was measured with a SQUID magnetometer (MPMS-XL5, Quantum Design Inc.) at 0.1 T.

\section{Results and Discussion}

\begin{figure}[h]
\includegraphics[width=88mm]{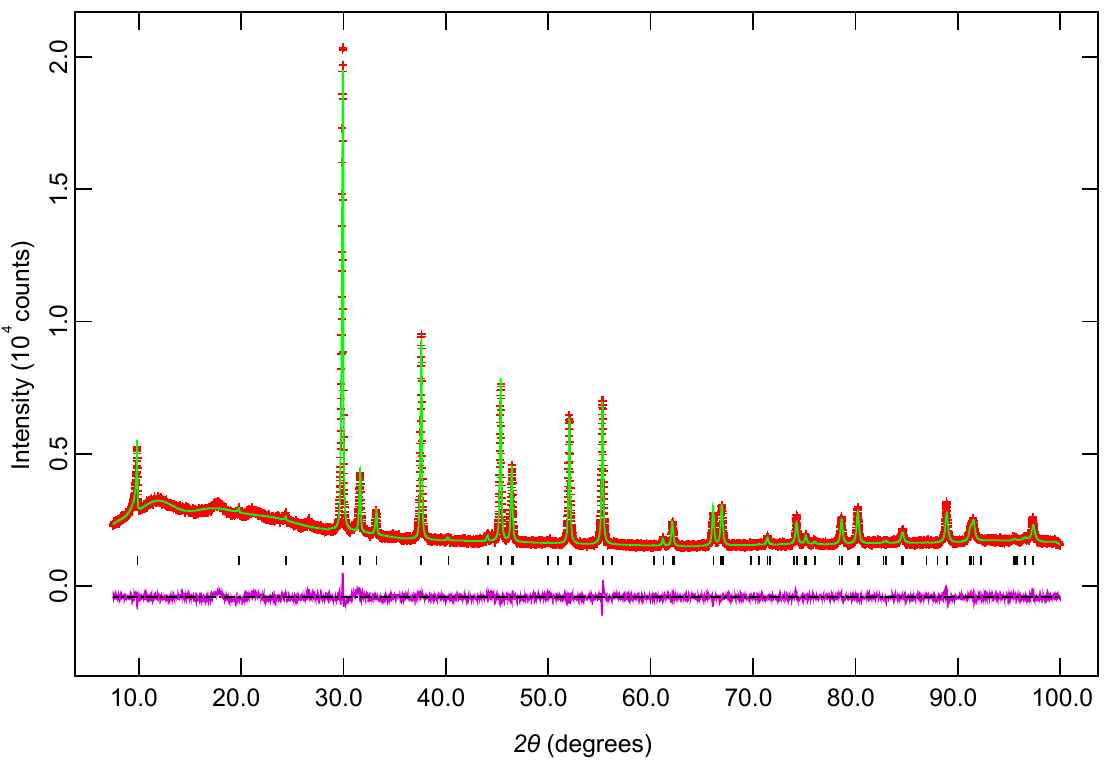}
\caption{(Color online) X-ray powder pattern (+) and Rietveld fit (-) of \SFA~at 297~K (space group $P4/nmm$).}
\label{fig:Rietveld_297}
\end{figure}

\begin{figure}[h]
\includegraphics[width=88mm]{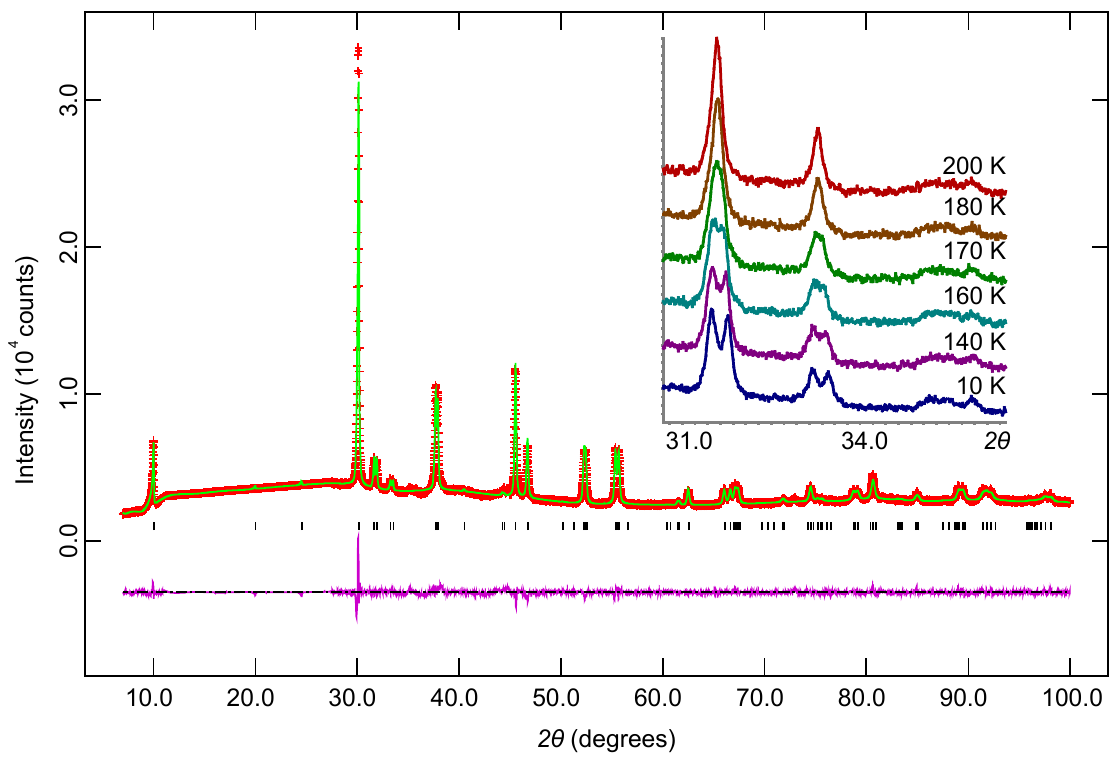}
\caption{(Color online) X-ray powder pattern (+) and Rietveld fit (-) of \SFA~at 10~K (space group $Cmme$). For more accurate results, the range between 11 and 27.5 2$\theta$ was excluded from the Rietveld refinement. Inset: Splitting of the $(2~0~0)$ and $(2~0~1)$ reflections.}
\label{fig:Rietveld_10}
\end{figure}

Figure~\ref{fig:Rietveld_297} shows the X-ray powder pattern of \SFA, which was completely fitted with a single tetragonal ZrCuSiAs-type SrFeAsF phase. Impurities are, if at all, less than 1~\%. In order to check for a structural phase transition, we have subsequently recorded patterns between 297 and 10~K. Several reflections broaden below 185~K and clearly split with further decreasing temperature. Patterns below 185~K were indexed with an orthorhombic $C$-centered unit cell ($a_{ortho} = \sqrt{2} \cdot a_{tetra}-\delta$; $b_{ortho} = \sqrt{2}\cdot b_{tetra}+\delta$; $c_{ortho} \approx c_{tetra}$; $\delta \approx$ 2.2 pm). The low temperature data could be refined in the space group $Cmme$. Figure~\ref{fig:Rietveld_10} shows the Rietveld fit of the data at 10~K and the continuous transition of the $(2~0~0)$ and $(2~0~1)$ reflections in the inset.

\begin{figure}[h]
\includegraphics[width=88mm]{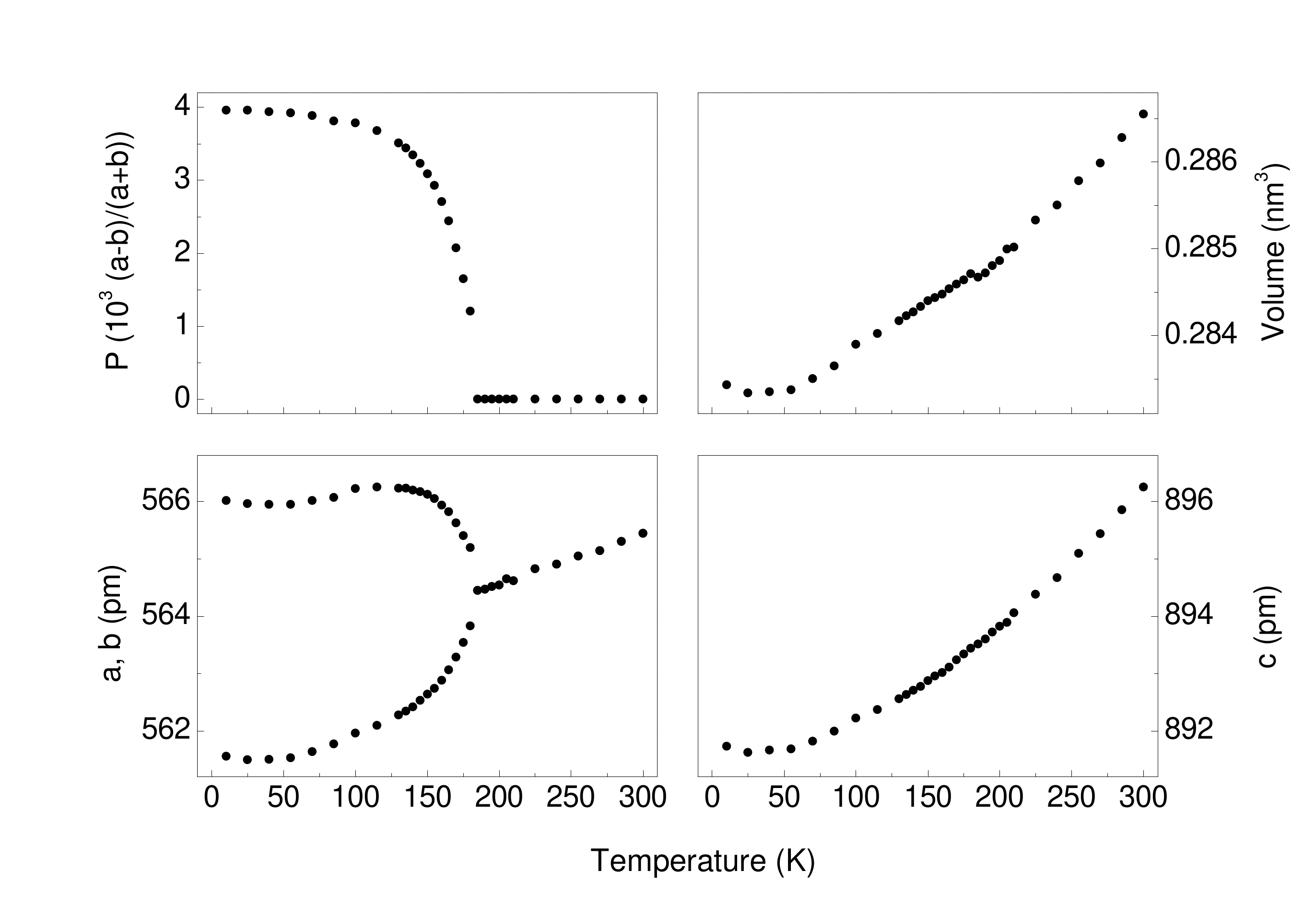}
\caption{Variation of the lattice parameters with temperature and order parameter $P = \frac{a-b}{a+b}$. The temperature dependence of the order parameter does not follow a simple power law. Values a and b for the tetragonal phase above 180~K are multiplied by $\sqrt{2}$ for comparability. Error bars are within data points.}
\label{fig:Lattice}
\end{figure}

The variation of the lattice parameters and the temperature dependence of the order parameter $P = \frac{a-b}{a+b}$ is depicted in Figure~\ref{fig:Lattice}. The $a$ lattice parameter in the tetragonal structure has been multiplied by $\sqrt2$ for comparison. We have found no signs of abrupt splitting of any reflections and also lattice parameters on cooling, but rather a continuous broadening. The space group $Cmme$ is a subgroup of $P4/nmm$ and therefore a second order phase transition is in agreement with our data. In terms of group theory, this transition is \textit{translationengleich} with index 2 ($P4/nmm~\rightarrow~Cmme$). Crystallographic data of SrFeAsF are summarized in Table~\ref{tab:Crystallographic}. The main effect of the phase transition appears in the Fe--Fe distances, where four equal bonds of 282.3 pm length split into two pairs of 283.0 and 280.7 pm length, respectively. This supports the idea, that the Fe--Fe interactions are strongly correlated with the SDW anomaly and may play a certain role for the properties of \SFA.

\begin{table}[p]
\caption{Crystallographic data of \SFA.}
\label{tab:Crystallographic}
\begin{center}
\begin{tabular}{lll}
 Temperature (K) & 297    & 10 \\
 Space group & $P4/nmm$           & $Cmme$ \\
 \textit{a} (pm) & 399.30(1)        & 561.55(1) \\
 \textit{b} (pm) & $=a$             & 566.02(1) \\
 \textit{c} (pm) & 895.46(1)       & 891.73(2) \\
 \textit{V} (nm$^{3}$) & 0.14277(1) & 0.28343(1) \\
 \textit{Z} & 2                   & 4\\
 data points & 9250               & 7649 \\
 reflections & 66                 & 97 \\
 atomic variables & 6            & 6 \\
 profile variables & 4           & 4 \\
 \textit {d} range & $0.992 - 8.955$ & $0.988 - 8.917$ \\
 R$_P$, \textit{w}R$_P$ & 0.0311, 0.0408 & 0.0385, 0.0611\\
 R$(F^{2})$, $\chi^{2}$ & 0.0408, 1.771 & 0.0552, 2.860\\
Atomic parameters: \\
 Sr & 2$c$ ($\frac{1}{4},\frac{1}{4},z$)                         &  4$g$ ($0,\frac{1}{4},z$)\\
    & $z$ = 0.1598(2)                    &  $z$ = 0.1635(2) \\
    & $U_{iso} = 185(9)$                    & $U_{iso} = 227(12)$            \\
 Fe & 2$b$ ($\frac{3}{2},\frac{1}{4},\frac{1}{2}$)   &  4$b$ ($\frac{1}{4},0,\frac{1}{2}$)\\
    & $U_{iso} = 75(8)$                   & $U_{iso} = 142(12)$\\
 As & 2$c$ ($\frac{1}{4},\frac{1}{4},z$)                       &  4$g$ ($0,\frac{1}{4},z$)  \\
    & $z$ = 0.6527(2)                    &  $z$ = 0.6494(2) \\
    & $U_{iso} = 46(8)$                  & $U_{iso} = 60(11)$\\
 F & 2$a$ ($\frac{3}{4},\frac{1}{4},0$)   &  4$a$ ($\frac{1}{4},0,0$)\\
    & $U_{iso} = 210(26)$               & $U_{iso} = 114(33)$\\
Bond lengths (pm):\\
Sr--F  &  245.7(1)$\times$4          & 247.0(1)$\times$4\\
Fe--As  &  242.0(1)$\times$4          & 239.7(1)$\times$4\\
Fe--Fe  &  282.3(1)$\times$4          & 283.0(1)$\times$2, \\
           &                     &    280.7(1)$\times$2\\
Bond angles (deg):\\
As--Fe--As &  111.2(1)$\times$2       & 112.5(1)$\times$2,\\
           &  108.6(1)$\times$4       & 108.3(1)$\times$2,\\
           &                     &    107.7(1)$\times$2\\
Sr--F--Sr & 108.7(1)$\times$2       & 110.7(1)$\times$2,\\
           &  109.8(1)$\times$4       & 110.1(1)$\times$2,\\
           &                     &    107.6(1)$\times$2

\end{tabular}
\end{center}
\end{table}

%
%

\begin{figure}[h]
\includegraphics[width=88mm]{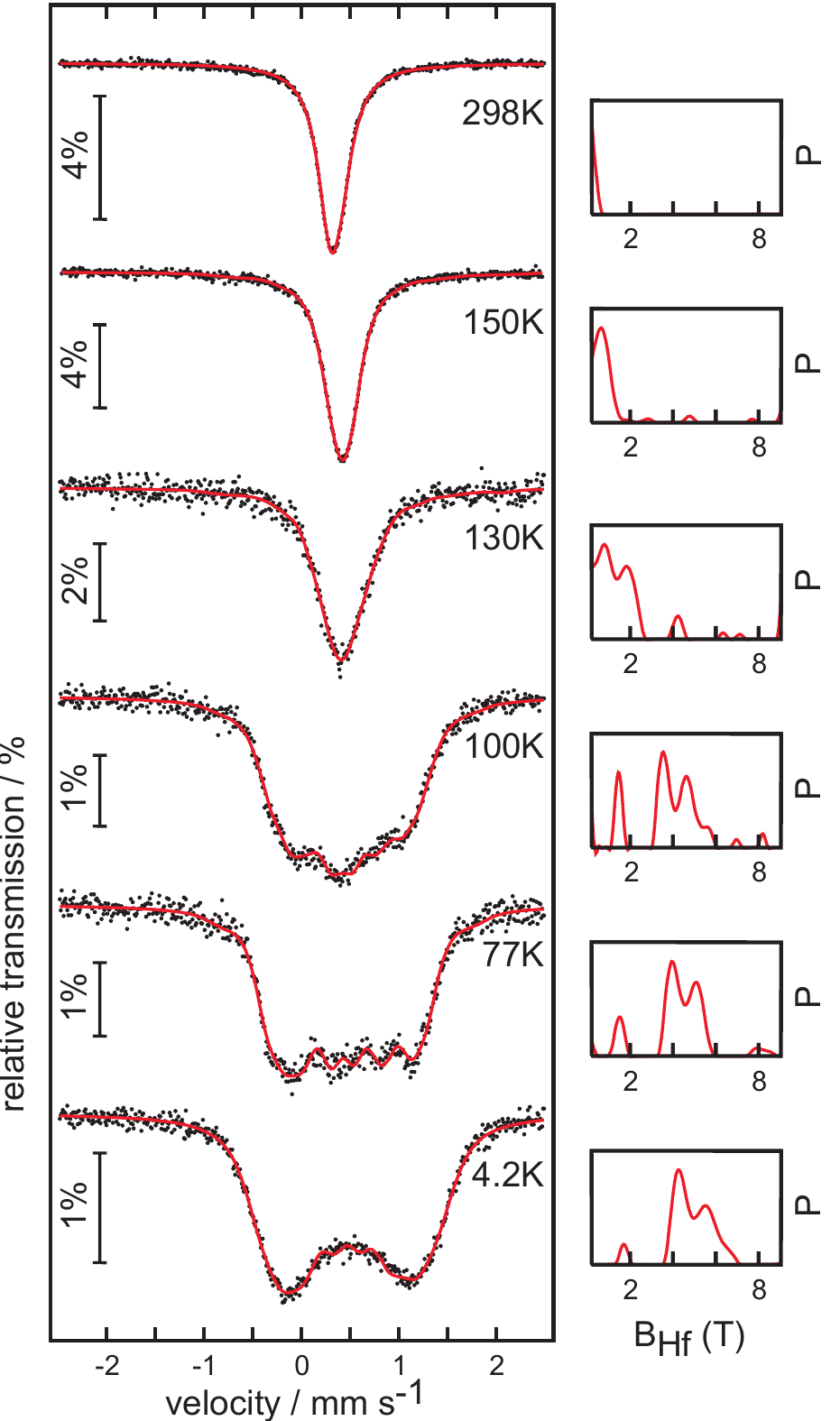}
\caption{Experimental and simulated $^{57}$Fe M\"{o}ssbauer spectra of \SFA. The temperature evolution of the field distribution functions, $P(B_{hf})$, obtained from computer fits at various temperatures are presented at the right-hand parts of the spectra. For details see text.}
\label{fig:Moessbauer}
\end{figure}

In order to check if the structural transition of SrFeAsF is connected with magnetic ordering and thus consistent with a SDW anomaly, we have recorded $^{57}$Fe M\"{o}ssbauer spectra at various temperatures, which are shown in Figure~\ref{fig:Moessbauer} together with transmission integral fits. The corresponding fitting parameters are listed in Table~\ref{tab:moessbauer}. In agreement with the crystal structure, at room temperature we observed a single signal at an isomer shift of $\delta$ = 0.33(1)~mm/s and an experimental line width $\Gamma$ = 0.30(1)~mm/s subject to quadrupole splitting of $\Delta E_{Q}$ =~--0.11(1)~mm/s. The non-cubic site symmetry of the iron atoms is reflected in the quadrupole splitting value. These parameters compare well with the recently reported $^{57}$Fe data for LaFeAsO\cite{Moessbauer_LaFeAsO-A,Moessbauer_LaFeAsO-B}, LaFePO\cite{Moessbauer_LaFePO}, \SrFA \cite{Moessbauer_SrFe2As2}, and \BaFA.\cite{Rotter-1-2008}
Below the SDW transition temperature, we observed line broadening of the signal followed by hyperfine field splitting due to the onset of magnetic order. As it can be clearly seen, there is a large evolution of the shape of the spectra with decreasing temperature. In the magnetically ordered state, the $^{57}$Fe M\"{o}ssbauer spectra of the \SFA~sample show significant differences when compared to the related systems \SrFA \cite{Moessbauer_SrFe2As2} and \BaFA. \cite{Rotter-1-2008} It was not possible to reliably analyze the data with only one magnetically split component below the SDW transition temperature. However, the spectra could be effectively fitted assuming a distribution of hyperfine magnetic fields (multi-component analysis), an approach similar to the WIVEL and M\O RUP method. \cite{Moessbauer_Wivel} The probability distribution curves $P(B_{hf})$ as a function of temperature are shown in Figure~\ref{fig:Moessbauer}. As a test for the program system we have also fitted the \BaFA~data from ref.~\cite{Rotter-1-2008} with the same algorithm. Since \BaFA~shows almost full hyperfine field splitting at 77~K, the distribution curve shows a single sharp contribution around 5~T.
For \SFA~we observed a shift of the fled components towards higher hyperfine fields with decreasing temperature. During numerical analysis, special attention was paid to the line width and quadrupole splitting parameters. These values were held constant at the values listed in Table~\ref{tab:moessbauer}. The magnitudes of the average hyperfine fields obtained from the data analysis are also listed in Table~\ref{tab:moessbauer}.

\begin{table}
\caption{Fitting parameters of $^{57}$Fe M\"{o}ssbauer spectra of \SFA~at different temperartures. $\delta$: isomer shift; $\Delta E_{Q}$: electric quadrupole splitting parameter; $\Gamma$: experimental line width; $B_{hf}$: average hyperfine field at the iron nuclei (for details see text). Parameters marked with an asterisk were held constant during the fitting procedure.  $\Gamma$, $\delta$ and $\Delta E_{Q}$ are given in $(mm\cdot s^{-1})$.}
\label{tab:moessbauer}
\begin{center}
\begin{tabular}{lcccr}
 $T$ (K) & $\Gamma$ & $\delta$ & $\Delta E_{Q}$ & $B_{hf}$\\
298 & 0.30* &  0.32*  & 0.08(1)  & --\\
130 &0.26*  &  0.40*  & -0.01(1) &  1.78(1)\\
100 & 0.26* &  0.45*  & 0.03(1)  &  3.64(1)\\
77   & 0.26*  & 0.45*   & 0.32(1) &  4.08(1)\\
4.2  & 0.28*  & 0.40*   & 0.20(1) &  4.84(3)
\end{tabular}
\end{center}
\end{table}

\vspace{0.8cm}

\begin{figure}[h]
\includegraphics[width=88mm]{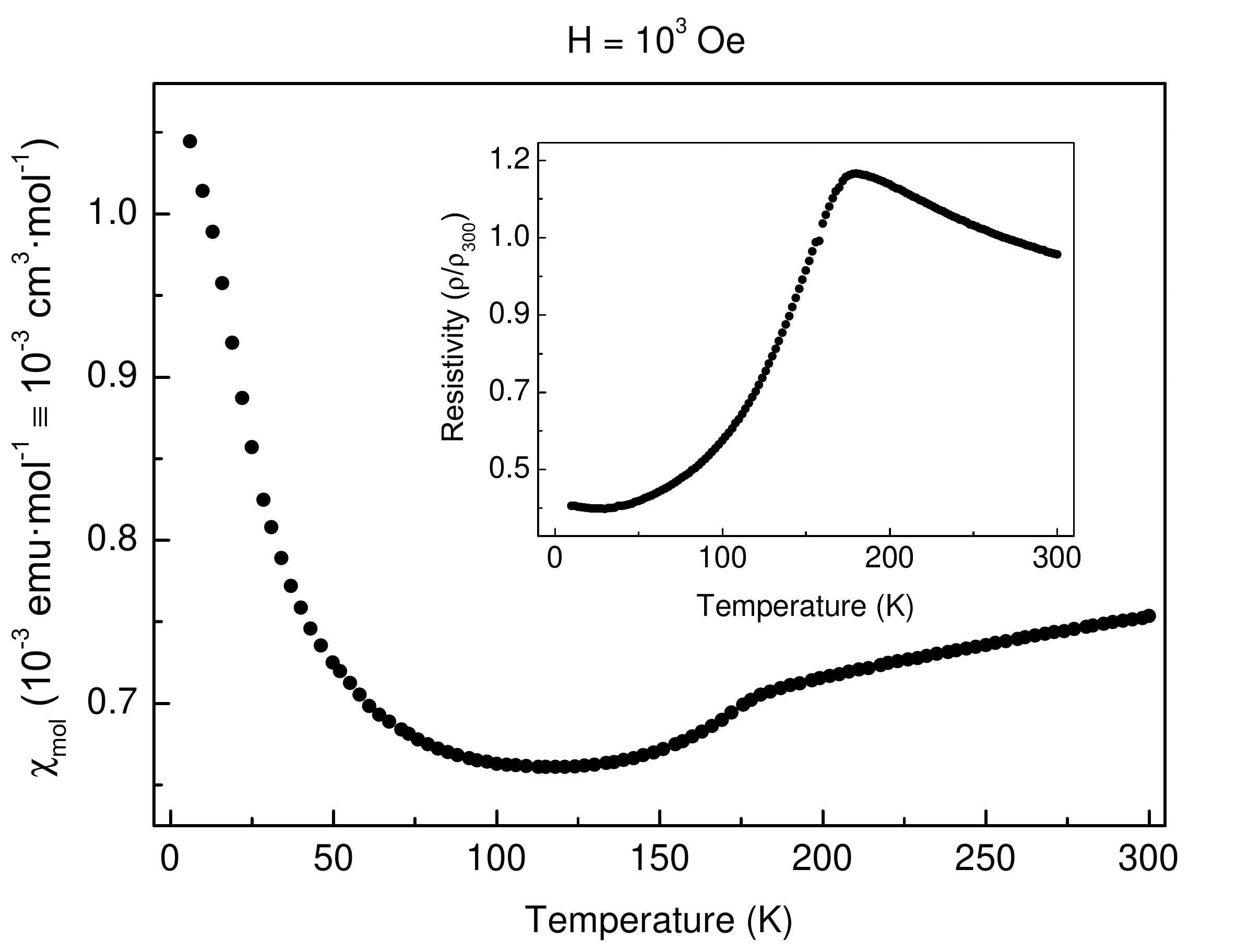}
\caption{Magnetic susceptibility of \SFA~measured at 0.1 T. Inset: $dc$ electrical resistance ($I = 100~\mu$A).}
\label{fig:chi_rho}
\end{figure}

So far, our results clearly prove a structural distortion in \SFA, connected with magnetic ordering. The nature of this effect is completely analogous to that in LaFeAsO. \cite{Nomura-2008} The transition is driven by a spin density (SDW) instability within the iron layers and therefore causes anomalies in the electrical resistivity and magnetic susceptibility. We have measured these properties of \SFA~and found the same behavior. The temperature dependencies of the magnetic susceptibility and $dc$ electrical resistance are depicted in Figure~\ref{fig:chi_rho}. \SFA~shows a relatively high resistance at room temperature. At 175~K, the resistance drops abruptly and then decreases slowly with lower temperatures. \SFA~shows a weak and only slightly temperature-dependent paramagnetism, as typical for a Pauli-paramagnetic metal. Below 175~K, $\chi$ drops at first but it increases again below 115~K.

%
%

\begin{figure}[h]
\center{
\includegraphics[width=80mm]{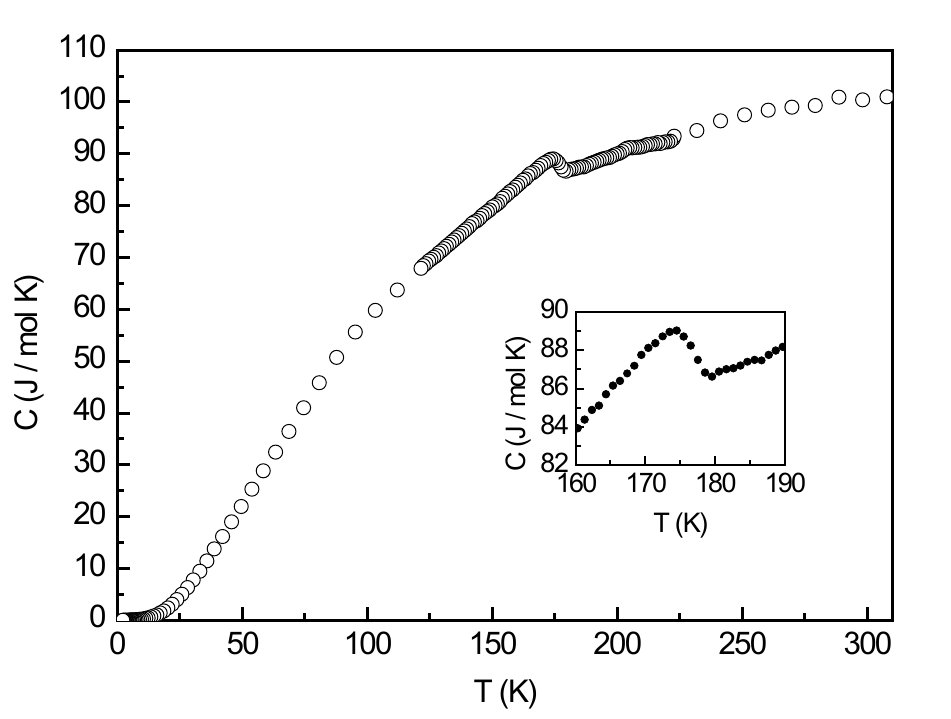}
\caption{Heat capacity ($C_{P}$) vs. temperature for SrFeAsF. The inset highlights the features between 160 and 190 K.}
\label{fig:SpecHeat}
}
\end{figure}

The specific heat data of SrFeAsF are shown in Figure~\ref{fig:SpecHeat}. At room temperature the specific heat capacity reaches the value of approximately 100 J/mol$\cdot$K, in agreement with the law of Dulong and Petit. The phase transition of SrFeAsF becomes noticeable by an anomaly, which can be seen in Figure~\ref{fig:SpecHeat}. The transition temperature has been determined as the inflection point of the anomaly (on the right-hand side). Thus, the $C_{p}$ data are in line with the resistivity, X-ray powder diffraction, and M\"{o}ssbauer spectroscopic data. In the low temperature region, the specific heat is of the form $C_p = \gamma T + \beta T^3$, where $\gamma$ is the electronic contribution and $\beta$ is the lattice contribution. The Debye temperature $\Theta_{D}$ can be estimated from the equation $\beta = (12\pi^4nk_B)/(5\Theta_D^3)$, where $k_B$ is the Boltzmann constant and $n$ is the number of atoms per formula unit. From the plot of $C_p / T$ vs. $T^2$ in the temperature range 4--10~K we found $\gamma$ = 1.5(2)~mJ$\cdot$K$^{-2}\cdot$mol$^{-1}$, $\beta$ = 0.2(1)~mJ$\cdot$K$^{-4}\cdot$mol$^{-1}$, and $\Theta_D$ =  339(1)~K.

\vspace{1cm}

%
%
\section{Summary}
We have synthesized the quaternary iron arsenide fluoride \SFA~, which crystallizes in the ZrCuSiAs-type structure. The properties of \SFA~are very similar to LaFeAsO, the parent compound of the recently discovered superconductors. Both materials are poor metals at room temperature and undergo second order structural and magnetic phase transitions. The $^{57}$Fe M\"ossbauer data of \SFA~show hyperfine field splitting, which indicates antiferromagnetic ordering occurring below the structural transition temperature. Consequently, \SFA~exhibits the same SDW anomaly at 175~K as LaFeAsO at 150~K. Since this SDW instability is known to be an important prerequisite for high-$T_C$ superconductivity in iron arsenides, our results strongly suggest that \SFA~can serve as a parent compound for a new, oxygen-free class of iron arsenide superconductors with ZrCuSiAs-type structure. It is highly probable that superconductivity in \SFA~can be induced either by electron or hole doping.

While we were finalizing our manuscript, we took notice of a recently published article on Co-doped CaFeAsF.\cite{Hosono-CaFeAsF} Therein, superconductivity was found below 22~K, which renders this new class of materials a very promising candidate for superconductors with high critical temperatures.

%
%

\acknowledgments
This work was financially supported by the DFG.

%
%

\end{document}